\documentclass[aps,prl,showpacs,twocolumn,a4paper,unsortedaddress,amsmath,amssymb,byrevtex]{revtex4}

\usepackage[latin1]{inputenc}
\usepackage{graphicx}
\usepackage{color}
\usepackage{dcolumn}
\usepackage{bm}
\usepackage{hyperref}
\usepackage{times}

\begin{document}

\title{Huge anisotropic magneto-resistance in iridium atomic chains}

\author{V. M. Garc\'{\i}a-Su\'arez$^2$}
\author{D. Zs. Manrique$^{1,2}$}
\author{C. J. Lambert$^2$}
\author{J. Ferrer$^{1,*}$}

\affiliation{$^1$  Departamento de F\'{\i}sica, Universidad de Oviedo, 33007 Oviedo, Spain}
\affiliation{$^2$ Departament of Physics, Lancaster University, Lancaster LA1
4YB, UK}
\affiliation{$^*$ e-mail: ferrer@condmat.uniovi.es.}
\date{\today}

\begin{abstract}
We analyze in this article the magneto-resistance ratio of finite
and infinite iridium and platinum chains. Our calculations, that
are based on a combination of non equilibrium Green function
techniques and density functional theory, include a fully
self-consistent treatment of non-collinear magnetism and of the
spin-orbit interaction. They  indicate that, in addition to having
an extremely large magnetic anisotropy that may overcome the super-paramagnetic
limit, infinite and also
realistic finite-length iridium chains show sizeable anisotropic
magnetoresistance ratios. We therefore propose iridium
nanostructures as promising candidates for nanospintronics logic
devices.
\end{abstract}

\pacs{85.75.-d,73.63.Nm,75.30.Gw}

\maketitle

The ability to enhance and tailor at the same time the magnetic anisotropy and
magneto-resistance of atomic-sized magnetic bits and junctions
will determine whether nanospintronics will be a viable
technology. Cobalt clusters and chains with magnetic
anisotropy barriers of the order of 1 meV were fabricated a few 
years ago by deposition on substrates of 5d elements\cite{gambardella1,gambardella2}.
Furthermore, several
recent theoretical predictions have pointed out that atomic
clusters and chains made of 4d elements should have even higher
magnetic anisotropy barriers than Co nanostructures\cite{blugel06,so-clusters,Carlo}. 
In the case of 5d
elements like platinum and iridium, calculations predict that the
anisotropies of iridium and platinum atomic clusters and chains
can be even larger than room-temperature thermal energies, so that
these nanostructures can overcome the technologically relevant
superparamagnetic limit\cite{so-clusters,predictions,Tos08}. 

Suspended chains of gold atoms connecting two gold electrodes were 
fabricated using the scanning tunneling microscope
and mechanically controllable break junction (MCBJ)
techniques\cite{ohn,yan}, where a quantized conductance close to
$G_0=2\,e^2/h$ was measured. Subsequent experiments found that
platinum and iridium also form suspended
chains, but that other elements like Ni, Co, Pd or Rh  do not
\cite{smit, Smit-thesis}, unless doped with oxygen or other
elements\cite{thijssen}. It was also demonstrated that longer
atomic chains could be made by deposition on stepped
surfaces\cite{gambardella1}, or by encapsulation in carbon
nanotubes\cite{mildred}.

The reduced dimensionality of nanometric objects increases their
tendency towards magnetism. Ugarte and coworkers proposed that
their experimental conductance data for platinum chains could be
explained provided that these were magnetic\cite{ugarte}.
Subsequent theoretical work indeed found that 4d and 5d infinite
linear chains of bulk-paramagnetic materials do become
magnetic\cite{Delin1,Delin2}. Very recently, several groups have
found that the magnetic anisotropy energy (MAE) per atom of
infinite zigzag platinum and iridium chains\cite{predictions}, and
of linear platinum chains\cite{Tos08} show values as large as 50
meV to 100 meV when stretched, which correspond to equivalent
temperatures of 500K to 1000 K. This suggests that the blocking
temperature $T_B$ of such nanostructures will attain a similar
value and therefore they should overcome the
superparamagnetic limit, so that their magnetization would be
finite at room temperature.

We draw the attention in this article to iridium and
platinum atomic structures as promising candidates for
nanospintronics. We show that iridium atomic chains not only
have huge magnetic anisotropies, but also meet the
requirement of having a significant magneto-resistance ratio.
The simulations presented here are based on density functional
theory\cite{Koh65} as implemented in the SIESTA code\cite{SIESTA},
and include a fully self-consistent and spin non-collinear
implementation of the spin-orbit interaction\cite{Lucas}. This
spin-orbit implementation has been successfully used to compute
the magnetic anisotropy of infinite Ir and Pt atomic
chains\cite{predictions} and of transition metal atomic
clusters\cite{so-clusters}, in agreement with the results of other
simulations\cite{blugel06,Carlo,Tos08}. The conductance
calculations were carried out with our spin non-collinear code
SMEAGOL\cite{Roc06}, which includes the fully self-consistent 
implementation of the spin-orbit interaction referred above.
We note that SMEAGOL has been
shown to provide conductance data that compare rather accurately
with the available experimental data on platinum
chains\cite{Gar05,Smit-thesis,smit}. The simulations have used the
Local Density Approximation\cite{Per81} and norm-conserving
pseudopotentials\cite{Tro91}. The atomic parameters that were fed
in the pseudopotential generator, as well as the basis set used
are similar to those reported in references
\cite{Gar05,predictions,so-clusters}.

Figure 1 (c) shows a schematic view of the experimental setup of a
MCBJ experiment, where at the last stages before breakage of a Pt
or Ir strip, a short suspended atomic chain is formed. In these
experiments, an electric current is applied to the circuit, and
the conductance $G$ is measured many times as the chain elongates.
The whole setup can also be understood as a magnetic junction that connects 
two paramagnetic electrodes. Recent work has shown that theory can  
reproduce accurately the experimental conductance data of finite-length platinum
chains\cite{Gar05}, but has not found significant differences
between the paramagnetic and the magnetic cases\cite{Pal05}. 

We show the energy of linear platinum and iridium chains as a function of $d_z$
in Figs. 2(a) and 3(a). 
It is important to notice that these chains are metastable. Indeed, allowing for 
lateral displacements of the atoms yields the energy curves in Figs. 2(b) and 3(b),
that only have minima for ladder or zigzag arrangements similar to those shown 
in Figs. 1(a) and (b). Furthermore, relaxation of the forces always finishes in 
zigzag or ladder chains, unless the atoms are constrained to lie in a linear array.
Actually, ladders are energetically more stable than zigzag geometries. But the relaxation
of the forces in a MCBJ geometry, like the one shown in Fig. 1(c) only leads to zigzag
arrangements, that gradually straighten as the electrodes are pulled apart.
Related to this effect, we note that infinite platinum (iridium) zigzag chains become linear only  
for distances $d_z$ longer than 2.55 \AA~ (2.40 \AA), since at these
distances, the zigzag angles fall abruptly to zero\cite{predictions}. These values of
$d_z$ are marked by a dotted vertical line in the right panels of
Figures 2 and 3. We will therefore only present results for the conductance of
infinite zigzag and linear, as well as finite zigzag, chains in this article. 

\begin{figure}
\includegraphics[width=0.7\columnwidth]{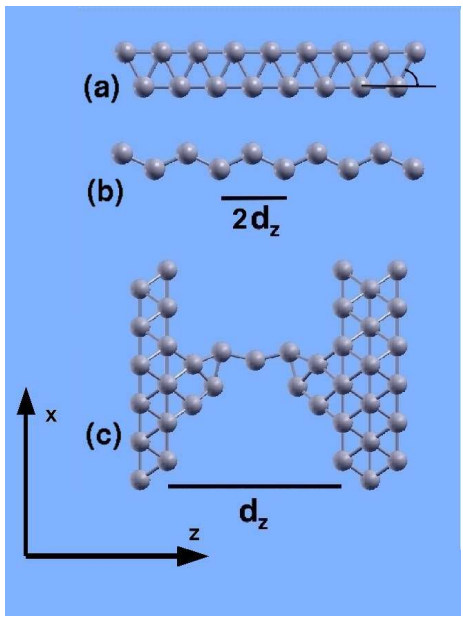}
\caption{(Color online) The geometry of the chains simulated in this
article, referred to the three coordinate axes. (a) Ladder
geometry that corresponds to the absolute minimum of the infinite
chains, see the top right panel in Figures 2 and 3; (b) zigzag
geometry that corresponds to the second minimum of the infinite
chains, see the top right panels in Figures 2 and 3; (c) schematic
view of the geometry of the relaxed finite-length chains in
contact to two electrodes. }
\end{figure}

\begin{figure}
\includegraphics[width=\columnwidth]{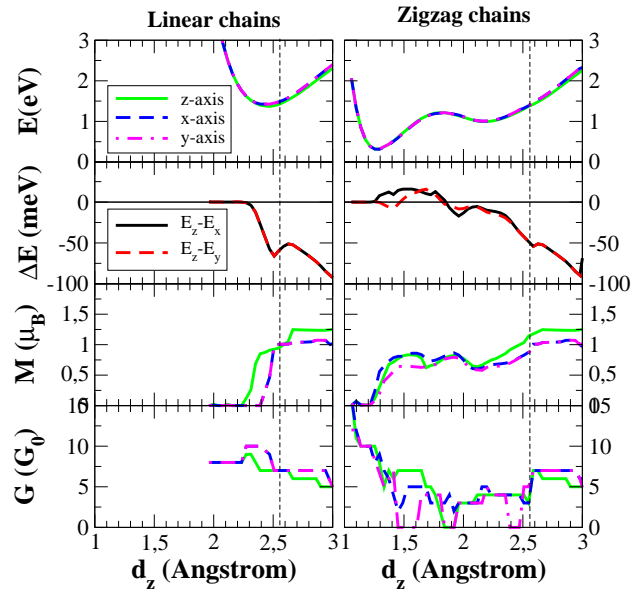}
\caption{(color online) Results of the simulations of infinite platinum
chains as a function of the distance $d_z$ defined in Figure 1 for
orientations of the spins along the $=x$, $y$ and $z$ axes. 
The left panels correspond to the case of metastable linear
chains. The
right panels corresponds to the geometries (b) and (c) in Figure 1.
The dashed vertical line indicates the
distance $d_z$ at which the zigzag chains cease to be stable.
(a) Energies $E_{x,y,z}$ per atom; (b) magnetic anisotropy
energies per atom $\Delta_{x,y}$; (c) spin moments per atom; (d) conductance. }
\end{figure}

We have performed separate
simulations for orientations of the atomic spins along the chain
axis ($z$- or easy-axis), perpendicular to it, but still in the
zigzag plane ($x$-axis) and perpendicular to the zigzag plane
($y$-axis), but the energies $E_{x,y,z}$  can not be discriminated in the
energy scale used in Figs. 2(a,b) and 3(a,b). We have therefore
computed\cite{predictions} the energy differences
$\Delta_{x,y}=E_z-E_{x,y}$ per atom, that we define as the MAEs, and
plotted them in Figs. 2(c,d) and 3(c,d) for platinum and iridium 
respectively. Our definition implies that when both $\Delta_{x,y}$ are 
negative, the magnetization lies parallel to the easy axis. On the contrary,
when $\Delta_{x,y}$ is positive, the magnetization is oriented perpendicular to the
easy axis, and aligns either parallel to the $x$- or the $y$-axes,
depending on whether $\Delta_x$ is larger or smaller than $\Delta_y$, respectively.
Notice that our results for linear
platinum chains (Fig. 2(c)) are similar to those presented in Ref.
\cite{Tos08}. More interestingly, the anisotropy of zigzag
platinum and iridium chains is $x$-axis for short $d_z$, but
shifts to easy axis if the chains are elongated, with the MAE
achieving values as large as 50-80 meV depending on the specific
value of $d_z$. The third row in Figs. 2 and 3 shows the spin
moment per atom as a function of $d_z$. Smogunov and coworkers\cite{Tos08}
found that linear platinum chains become magnetized at a shorter elongation
$d_z$ for spin lying in the axis of the chain than for spins
oriented perpendicular to it. This fact led them to predict a
colossal magnetic anisotropy for the window of elongations within
the onset of both magnetizations. However, we note here that the
magnetization of the more realistic zigzag chains is finite for
almost the full range of elongations regardless of the spin
orientation, and therefore that effect is not expected to be seen. Interestingly,
we find that the magnetization of iridium chains decreases when
the chain is pulled apart, in a window of elongations around the equilibrium $d_z$.

\begin{figure}
\includegraphics[width=\columnwidth]{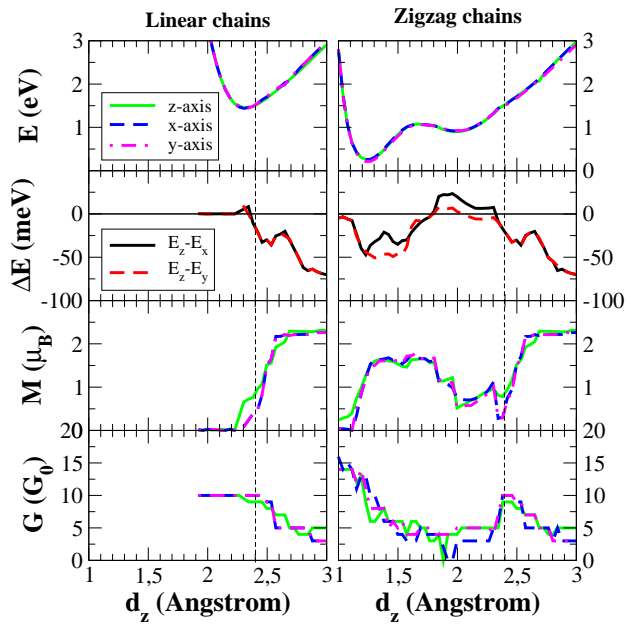}
\caption{(Color online) Results of the simulations of infinite iridium
chains as a function of the distance $d_z$ defined in Figure 1 for
orientations of the spins along the $=x$, $y$ and $z$ axes. 
The left panels correspond to the case of metastable linear
chains. The
right panels corresponds to the geometries (b) and (c) in Figure 1.
The dashed vertical line indicates the
distance $d_z$ at which the zigzag chains cease to be stable.
)a) Energies $E_{x,y,z}$ per atom; (b) magnetic anisotropy
energies per atom $\Delta_{x,y}$; (c) spin moments per atom; (d) conductance. }
\end{figure}

Since infinite Ir and Pt chains show large MAEs, an obvious
question is whether they also show a strong anisotropic
magneto-resistance effect that could be activated by the presence
of a magnetic field. If so, this effect could open the way towards
room-temperature inorganic nanomagnetic memory bits. We have
computed the low-voltage conductance $G_{x,y,z}$ for orientations
of the spins along the three axes, that we show in
Figs. 2(g,h) and 3(g,h), again as a function of $d_z$.
We have found that the magneto-resistance ratio, defined as
\begin{equation}
AMR_{x,y} = 2\,\frac{G_{x,y}-G_z}{G_{x,y}+G_z}
\end{equation}
\noindent is finite but small over wide ranges of elongations
$d_z$. Furthermore, iridium chains show
larger magneto-resistance ratios than platinum ones for a wider
range of $d_z$ at and above the equilibrium elongation. We therefore
expect that the magneto-resistance of iridium chains should be
more robust than that of platinum chains.

Remarkably, for certain windows of the elongations $d_z$ in zigzag chains,
one of the conductances may fall exactly to zero, marking the
onset of a perfect magneto-resistance ratio. Notice though  that these
windows of perfect magneto-resistivity do not appear for linear chains.
Zigzag platinum chains show three of those windows,
which occur when the atomic spin moments are oriented in the $y$-axis
(i.e. perpendicular to the zigzag plane). Two of these appear for
unrealistically short elongations $d_z$, but the third develops
for distances slightly larger that the equilibrium $d_z$ and
therefore could have the potential to be realized and measured
experimentally. Iridium chains show only one of these windows,
where $G_{x}$ vanishes, which occurs for elongations about the
equilibrium $d_z$, where the magnetic anisotropy favors
orientations of the spins along the $x$-axis. The MAE in this
window of elongations is of the order of 25 meV. Interestingly,
the vanishing conductance develops precisely for orientations of
the spins along the $x$-axis. 

Since the above finite magneto-resistance ratios originate from
the different number of channels available for electron conduction
close to the Fermi energy for the different spin orientations, we
now study the band structure of infinite chains. Indeed, the
number of conductance channels can be found by counting the number
of bands crossing the Fermi energy. We note that each band moves
up or down in energy at a different pace as the chain elongates.
Furthermore, since the spin-orbit interaction couples spin and
orbital degrees of freedom, these shifts are different for the
different spin orientations. Since the band structure of
non-conducting chains must show a mini-gap at the Fermi energy, we
have plotted in Fig. 4 the band structure of a platinum chain at
$d_z=2.45$ \AA~ for paramagnetic as well as for spin-orbit
simulations. We indeed find that when the atomic spins point in
the $y$-axis there exists a small mini-gap $E_g=25$ meV at
$k=0.4\,\pi/d_z$. This value allows us to estimate the minimum
number of atoms that a finite-sized platinum chain must have to
develop such a mini-gaps. Using the relationship $N\simeq
\hbar\,v_F / E_g\,d_z$, where we estimate the Fermi velocity at
this $k$-point as $\hbar\,v_F\simeq1000$ meV $d_z$, we find a
critical length of about 40 atoms.  We therefore expect that it
will be difficult to measure a significant magneto-resistance
ratio in platinum chains. We have indeed simulated 5-atoms long
platinum chains contacted to (001) platinum electrodes, in a
geometry similar to Fig. 1 (c) and found that the
magneto-resistance ratio was essentially zero.

\begin{figure}
\centerline{\includegraphics[angle=-90,width=\columnwidth]{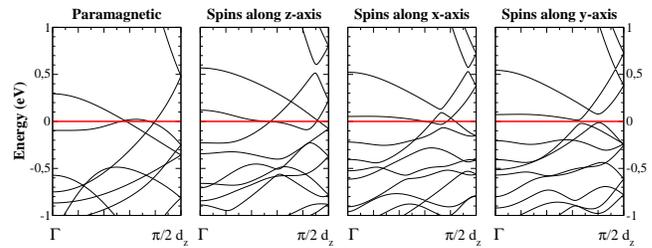}}
\caption{Band structure of a zigzag platinum chain with an
elongation $d_z=2.45$ \AA. From left to right the panels
show the bands for a paramagnetic calculation, and for spin-orbit
simulations whereby the spins are oriented along the $z$-, $x$-
and $y$-axes, respectively. }
\end{figure}

In contrast, Fig. 5 shows the band structures of iridium chains,
obtained from paramagnetic and also for spin-orbit simulations for an
elongation $d_z$ such that $G_x$ falls to zero. We note that the
spin-orbit interaction generates much larger gaps for iridium than
for platinum. For this particular elongation, the gap when the
spins point along the $x$-axis is positioned exactly at the Fermi
energy and has a value $E_g\simeq 100$ meV at the $\Gamma$ point.
Using the estimated Fermi velocity $\hbar\,v_F\simeq1500$ meV
$d_z$, we predict that chains containing  about 15 atoms should
show large magneto-resistance ratios.

\begin{figure}
\centerline{\includegraphics[angle=-90,width=\columnwidth]{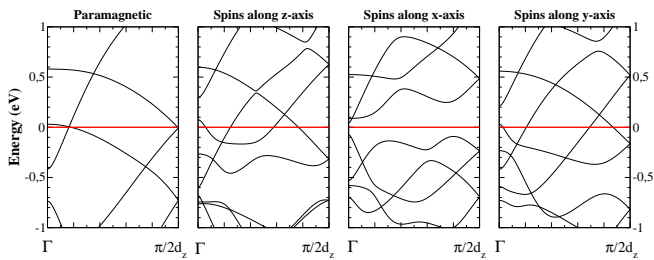}}
\caption{Band structure of a zigzag iridium chain with an
elongation $d_z=1.96$ \AA. From left to right the panels
show the bands for a paramagnetic calculation, and for spin-orbit
simulations whereby the spins are oriented along the $z$-, $x$-
and $y$-axes, respectively. }
\end{figure}

To confirm this, we have simulated 2- and 21-atoms long iridium
chains in contact to (111) iridium electrodes terminated by a
pyramid, as in Fig. 1 (c), for a number of elongations of the
chain, which cover both the equilibrium distance and also stretched
situations. We have found that the magneto-resistive ratio of the 2-atom
chains was negligible for whatever of these elongation. In contrast, 
we have found that $G_{x}$ is indeed substantially smaller than $G_{y,z}$, 
as we show in Fig. 6, rendering a finite ratio $AMR_x$. We stress that
our previous analysis of Figs. 3 and 5 indeed led us to expect that $G_x$
had to be smaller than $G_{y,z}$.

\begin{figure}[t]
\centerline{\includegraphics[angle=-90,width=0.8\columnwidth]{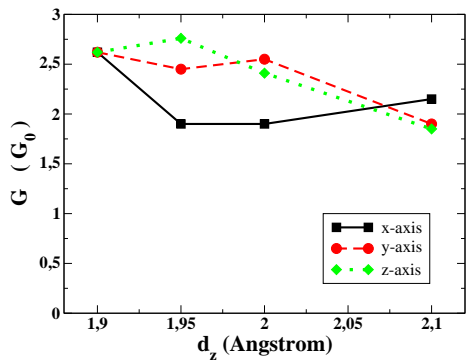}}
\caption{(Color online) Conductance as a function of $d_z$ of iridium chains 
containing 21 atoms. }
\end{figure}

To summarize, we expect that short platinum chains should show a
negligible magneto-resistance ratio until the number of atoms in
the chain exceeds of order 40. Since such a length is very
difficult to attain experimentally, we believe that platinum is
not the optimum magneto-resistance material. We predict that a
better candidate element is iridium. Short iridium chains are
expected to display small but measurable magneto-resistance
ratios. Longer chains could develop perfect magneto-resistance for
certain elongations. These clear differences are experimentally
accessible and therefore we anticipate that future experimental
studies of iridium chains will prove to be extremely fruitful.

We acknowledge useful conversations with C. Untiedt, J. J.
Palacios and A. Halbritter. L. 
Fern\'andez-Seivane tested the gaps in platinum
chains using the code Quantum ESPRESSO (http://www.pwscf.org). 
We acknowledge financial support
from the European Union (MRTN-CT-2003-504574), the
Spanish MEC (MEC-FIS2006-12117) and the UK EPSRC.

\end{document}